\documentstyle[12pt]{article}
\newcommand{\beq}{\begin{equation}}
\newcommand{\beqa}{\begin{eqnarray}}
\newcommand{\eeqa}{\end{eqnarray}}
\newcommand{\eeq}{\end{equation}}
\newcommand{\nn}{\nonumber}

\newcommand{\trd}{\mbox{$Tr_{D}$}}
\newcommand{\order}{\mbox{$\cal O$}}
\newcommand{\slashv}{\mbox{$\not v$}}

\newcommand{\chiral}{$SU_{L}(3)\times SU_{R}(3)$}
\newcommand{\sun}{$SU(3)$}

\newcommand{\gmu}{\mbox{$\gamma_{\mu}$}}
\newcommand{\gnu}{\mbox{$\gamma_{\nu}$}}
\newcommand{\grho}{\mbox{$\gamma_{\rho}$}}

\newcommand{\gfive}{\mbox{$\gamma_{5}$}}

\newcommand{\Hm}{\mbox {$\cal H$}}
\newcommand{\Hz}{{\cal H}}
\newcommand{\udag}{\mbox{$u^{\dagger}$}}
\newcommand{\dmu}{\mbox{$\partial_{\mu}$}}

\newcommand{\ra}{\rightarrow}

\newcommand{\np}{Nucl.\ Phys.\ B }
\newcommand{\pl}{Phys.\ Lett.\ }
\newcommand{\prd}{Phys.\ Rev.\ D }
\newcommand{\prl}{Phys.\ Rev.\ Lett.\ }

\newcommand{\zp}{Z.\ Phys.\ C }
\begin{document}
\baselineskip 18 pt plus 1pt minus 1pt
\begin{titlepage} ~\\
\hspace*{6 cm} CEBAF-TH-92-16
\begin{center}
{CHIRAL PERTURBATION THEORY FOR SU(3) BREAKING IN HEAVY MESON SYSTEMS }\\
\vspace*{1.cm} J. L. ${\rm Goity}^{\;\;\star}$ \footnotetext[1]{{
 $\star$}\
E-mail : goity@cebaf2.cebaf.gov}\\
 \vspace*{.6cm}       {\it  Continuous Electron Beam Accelerator Facility\\
                      Newport News, VA 23606}
\end{center}
 \vspace*{1.cm}

\begin{abstract}

The SU(3) breaking effects due to  light quark masses
 on heavy meson masses, decay constants
($F_{D}, F_{D_{s}}$) and the form factor for semileptonic
 $\overline{B}\rightarrow D^{(\ast)}
l\bar{\nu}_{l}$ transitions are formulated in chiral perturbation theory,
using a heavy meson effective Lagrangian and expanding in
inverse powers of the heavy meson mass.
To leading order in this expansion, the leading
chiral logarithms and the required counterterms are determined.
At this level, a
 non-analytic correction  to the mass splittings
of ${\cal O}(p^3)$ appears,
similar the the one found in light baryons.
The correction to $F_{D_{s}}/F_{D}$ is roughly estimated
 to be of the order of $10\%$
and, therefore, experimentally accessible, while the correction
to the form factor is likely to be substantially smaller.
We explicitly check that the heavy quark symmetry is preserved
by the chiral loops.

\end{abstract}
 \vspace*{0.6cm}
\begin{center}
May 1992
\end{center}
\end{titlepage}
\newpage
\section{Introduction}

In addition
to their intrinsic significance for the study of electroweak interactions
(quark mixing, rare decays, CP-violation),
heavy hadrons containing a single heavy quark $c$ or $b$
 might also prove to be
a useful tool for unveiling new aspects of the strong interactions.
The reason for this is the large approximate
symmetry available in these systems, which
constrains  the QCD dynamics and, therefore, substantially simplifies their
study. As the mass $m_{Q}$ of the heavy quark becomes much larger
than the characteristic QCD scale (say, $m_{\rho}$), in all strong
interaction processes where the relevant scale of momenta
is much smaller than the heavy quark mass, the heavy quark
approximately behaves as a static color source in the rest frame
of the hadron, with its spin
dynamically decoupled. In this limit, the velocity of the heavy quark
and its spin become conserved observables. This results in a
 superselection rule for
the velocity $\cite{E&H,georgi1}$, and a spin-flavor symmetry
 (the Isgur-Wise (I-W) symmetry) $\cite{IW1,V&S}$ which enjoys
all bonafide properties of an internal symmetry. In addition, the
light quark degrees of freedom in the heavy hadron carry information
about the chiral \chiral\ symmetry of QCD. In particular, chiral symmetry
dictates the form of the couplings between the heavy mesons and
the Goldstone bosons ($\pi,\;K,\;\eta$) resulting from the
Nambu-Goldstone nature of its realization (see for instance $\cite{goity}$).

The corrections to the symmetry limit are naturally obtained by
expanding in powers of $1/m_{Q}$ (more precisely in the present
 context, in powers of
$1/M_{H}$, where $M_{H}$ is the heavy hadron mass) and by treating the light
quark masses, which explicitly break chiral invariance, as
a perturbation. The expansion in powers of  $1/M_{H}$
has the virtue of enabling a systematic chiral expansion,
at each order in
 $1/M_{H}$, where the chiral power counting  is in correspondence
with  an expansion
in loops $\cite{J&M}$, similarly to chiral perturbation theory
in light mesons $\cite{g&l}$. This only applies to processes
involving only one heavy hadron, otherwise, infrared divergences
modify the naive chiral counting $\cite{weinberg}$.

\sun\  breaking effects induced by the light quark masses
are inherently of low energy character, and therefore,
suited to a systematic study within the chiral expansion. Recently,
various groups $\cite{Wise,Wiseetal,Burdman}$ have initiated this field, in
which
interesting theoretical results are expected to emerge$^{\S}$
 \footnotetext[1]{{ \S}
As we  became aware of the works in ref.
$\cite{Wise,Wiseetal,Burdman}$  the present project was approaching its
conclusion.
Some overlap with some of these references has been unavoidable.}.
 At present, \sun\  breaking
 is only observed through the mass splittings in
$D$, $D^{\ast}$ and $B$ mesons and in charmed baryons. In the future,
one  also expects observation through other quantities
(e.g. decay constants). This requires, however,
substantial improvement in strange D meson measurements.
As for the decay constants, at present only an upper bound
exists for non-strange D mesons: $F_{D} \leq 200 MeV$ ($F_{\pi}=93 MeV$).
For $B$ mesons, observation of \sun\   breaking in
 form factors (e.g. in $B_{(s)}\ra\overline{D}l\nu$ decays)
 could only be achieved in a B-meson factory. The corrections
to $B^{0}-\overline{B}^{0}$ and
$B_{s}-\overline{B}_{s}$  mixing are also of great interest, and have recently
been
analyzed $\cite{Wiseetal}$. Applications in connection with lattice QCD
simulations
of heavy-light systems can also be envisaged. For instance, finite volume
effects in the continuum limit,
 of relevance in this context, can be unambiguously determined
using the chiral expansion $\cite{goity}$.

 The predictive power of the chiral expansion is limited
by the counterterms which must be added at each stage.
The counterterms are ordered according to chiral power counting
and are required as subtractions to U.V. divergent chiral loops.
 To overcome this
drawback, a sufficient number of measured observables is needed
as input. While this is possible in light mesons, it is
not yet clear that it can be achieved in heavy mesons.

In this paper, we study the \sun\ breaking corrections, in the limit of
infinite heavy quark mass,
to the ratio  of decay constants
$F_{H_{s}}/F_{H_{u,d}}$ ($H$ denotes a heavy
meson), mass splittings, and the Isgur-Wise form factor associated with
the charged current in the transitions $B^{0}\ra D^{-}$ and
$B_{s}\ra \overline{D}_{s}$. The chiral logarithms and their associated
counterterms are determined to one-chiral loop order. The presence
of non-analytic contributions in the light quark masses
 $(\alpha \;m_{q}^{3/2})$
to the mass splittings is noticed.

\section{Effective Theory}

In this section, we discuss in detail a formulation of an effective
theory for heavy ($D,\;B$) mesons coupled to the Goldstone bosons of
spontaneously broken chiral symmetry. In this formulation, \sun\   breaking
effects can be consistently studied in a chiral-loop expansion. This
is made possible by simultaneously performing an expansion in powers
of the inverse of the  heavy meson mass.

The form of the interactions between Goldstone bosons
and heavy mesons containing
a heavy antiquark and a light quark are determined by the transformation
properties of the heavy meson wave functions under chiral \chiral.
The transformation law is easily found by the method of   Coleman, Wess and
Zumino
$\cite{callan}$. In our case, pseudoscalar and vector
 heavy mesons appear in triplets under flavor \sun,
and this fixes the transformation law under an arbitrary chiral transformation
$g=L\otimes R$ to have the following form:
\beq
g:H=h H~~~~~~~~,
\eeq
where $H$ denotes the heavy meson wave function and $h$ is a  $3\times 3$
\sun\
 matrix which depends on the octet of Goldstone
excitations. Although the explicit form of $h$ will not be  needed,
it can be determined
in the following manner: one defines a $3\times3$ \sun\  matrix $U(x)$
 parametrized
by the classical Goldstone fields, and whose transformation
law is given by $g:U(x)=L \;U(x)\; R^{\dagger}$,
 where on the  right
hand side ordinary matrix multiplication is meant. By means of
$u(x)\equiv\sqrt{U(x)}$, one can determine $h$ in such a way that
(1) is a realization of the chiral group. The dependence of
$h$  on the
Goldstone excitations  results from solving
the following system of equations:
\beqa
L\, u(x)&=& u'(x)\,h \nn \\
R\, u^{\dagger}(x)&=&  u'^{\dagger}(x)\,h
\eeqa
In what follows we will use the exponential parametrization for $u(x)$:
\beq
u(x)=exp\;(-i \frac{\pi_{a}(x)\lambda^{a}}{2 F_{0}})~~~~~~~~,
\eeq
where the  Goldstone fields $\pi_{a}(x)$
are real and identified with the light pseudoscalar octet  ($\pi,\;K,\;\eta$),
 $F_{0}\sim 93 MeV$ is
the pion decay constant in the chiral limit, and the Gell-Mann hermitian
matrices are normalized by $Tr(\lambda^{a}\lambda^{b})=2\delta^{ab}$.

In order to build the effective Lagrangian invariant under chiral
rotations, one needs to introduce a  covariant derivative under
the transformation law (1). Using the transformation properties
of $u(x)$ implied by (2), the explicit form of the covariant derivative
is given by:
\beqa
\nabla_{\mu}H&=&(\dmu+i\Gamma_{\mu}) H \nn \\
\Gamma_{\mu}=\Gamma^{\dagger}_{\mu}&=& \frac{i}{2}\;(u\dmu u^{\dagger}+
\udag\dmu u)
\eeqa
Besides the covariant derivative, the following hermitian pseudovector
must  also be considered:
\beq
\omega_{\mu}=\frac{i}{2}\;(u\dmu \udag - \udag \dmu u)
\eeq
whose transformation law under a chiral rotation is given by:
\mbox{$\;\;g:\omega_{\mu}=h \omega_{\mu} h^{\dagger}$}.

Having established the chiral transformation properties for the heavy mesons,
we now turn to the important
aspects connected with the large mass of the heavy quark.
As the mass of the heavy quark tends to infinity, the QCD Lagrangian
becomes invariant under a new global symmetry (I-W symmetry)
$\cite{IW1}$, which, in the case of a single heavy quark,
 corresponds to the
operation of independently  rotating the spins of the heavy quark and
antiquark. This
becomes an internal symmetry of QCD, which, for $N_{h}$ heavy quarks
is $SU(2 N_{h})\times SU(2N_{h})$ with one factor referring to quarks
and the other to antiquarks, as this two sectors become independent
in the infinite mass limit. The  lowest lying pseudoscalar and vector heavy
mesons,
relevant to this work,
belong to a multiplet under the I-W symmetry, and must therefore
be treated together.  Their masses are equal, up
to symmetry breaking corrections of hyperfine origin
equal to $\Lambda^{2}/m_{Q}$ ($\Lambda$ is a typical QCD scale, in
this specific case  $\Lambda\sim m_{\rho}/\sqrt{2}$)
, and, their
 transition amplitudes become  related.
For all amplitudes where the relevant scale is $\Lambda$, the heavy
quark velocity is conserved. These conservation laws  obviously extend
 to the heavy-meson Goldstone-boson
 interactions relevant in the present context. For this reason, it
is convenient to consider as the  starting point the effective QCD theory
for the heavy quark sector, defined in terms of  heavy quarks and antiquarks
separately (i.e., in the effective theory no virtual loops of
heavy quarks are required) for each 4-velocity $v_{\mu}$ $\cite{georgi1}$.
The corresponding effective theory for the heavy (anti)-mesons is
obtained in a similar
way, by defining non-relativistic fields for mesons ($\overline{Q}q$)
 and anti-mesons ($\bar{q}Q$) for a given 4-velocity $v_{\mu}$ as follows:
\beqa
H^{(+)}_{v}(x)&=&\sqrt{M_{H}}\;e^{i M_{H}v.x}\;\Psi_{+}(x)\nn\\
H^{(-)}_{v}(x)&=&\sqrt{M_{H}}\;e^{i M_{H}v.x}\;\Psi_{-}^{\dagger}(x)~~~~~~~,
\eeqa
where $\Psi_{+,-}(x)$ are the positive and negative frequency
components of the relativistic meson field $\Psi(x)$ and $M_{H}$ is the
heavy meson mass. Since the meson and antimeson sectors
become independent, we will only work with mesons, and the field
 $H^{(+)}_{v}(x)$,
which annihilates heavy mesons with velocity $v_{\mu}$,
 will be simply denoted by $H(x)$.

For heavy mesons, the I-W symmetry is elegantly implemented
by merging the pseudoscalar
 and vector mesons into a multiplet as follows $\cite{Falk,bjorken}$:
\beq
\Hz(x)=\frac{1+\slashv}{2}\;\left(-\gfive H(x)+\gmu H^{\mu}(x)\right)~~~~~~,
\eeq
where the vector field $H^{\mu}$ satisfies the constraint
$v_{\mu}H^{\mu}=0$. For further convenience, the field conjugated to
\Hm\, which will create heavy mesons,
 is defined by $\overline {{\cal H}}=\gamma_{0} \Hm ^{\dagger}\gamma_{0}$.
Under chiral operations \Hm\ transforms as indicated in (1), and under the
heavy quark symmetry operations its transformation law is the following:
\beqa
\Hz&\ra&e^{i \theta_{j}S_{j}}\;\Hz  \nn\\
S_{j}&=&i\;\epsilon_{jkl}\; [\;\;/\!\!\!e_{k}\;,\;\; /\!\!\!e_{l}\;]~~~~~~~,
\eeqa
where $e_{i}^{\mu},\; i=1,2,3$
are normalized space-like vectors orthogonal to $v_{\mu}$.

Since the definition (6) corresponds in the rest frame
of the meson to the subtraction of the rest mass
energy, the operator $-i\dmu$ acting on \Hm\ gives the residual
momentum. In particular, for the purposes of the chiral
expansion, this residual momentum will count as a quantity of
$\order(p)$. Similarly, $\dmu u$ is of $\order(p)$, and, therefore,
the covariant derivative  only contains terms of  $\order(p)$
as  is also the case for $\omega_{\mu}$.

It is now straightforward to write down the lowest order effective
Lagrangian, in
both chiral and $1/m_{Q}$ expansions, which is invariant under
chiral and I-W transformations as well as under parity and charge
conjugation. This Lagrangian is of $\order(p)$
and reads as follows:
\beqa
{\cal L}^{(1)}&=&-\frac{1}{2} \;v_{\mu}\; \trd \;\{\;\overline{\Hz}
\nabla^{\mu}\Hz\;\}
 \nn \\
&+&\frac{1}{2}\;g\; \trd\;\{\;\overline{\Hz}\;\omega^{\mu}\;\Hz\;\gmu\gfive\;\}
{}~~~~~~~,
\eeqa
where \trd\ denotes the trace over Dirac indices.
The first term contains the
kinetic energy and interactions of the heavy mesons with an even number of
Goldstone bosons and no change in the spin of the heavy meson.
This term is universal and automatically satisfies the I-W symmetry.
The second term gives rise to virtual transitions $H^{\ast}+m \pi
\leftrightarrow H+n\pi$
and  $H^{\ast}+m \pi\leftrightarrow
 H^{\ast}+n\pi$ with $(n+m)$ odd. The I-W symmetry
imposes that the strength of both types of transitions must be equal.
The corresponding
 adimensional coupling constant $g$ could be determined from the decay
\mbox{$D^{\ast +}\ra D+\pi$}; unfortunately, at present,  only
an upper bound
on the $D^{\ast +}$ width is available, resulting in
   $g^{2}\leq 4.8$. One may take as a rough estimate for
$g^{2}$ the
corresponding coupling constant in the K meson system: $g^{2}_{K^{\ast}K\pi}
=0.46$. The quark model result is $g^{2}\sim 0.3$ $\cite {IW3}$, while
 the QCD sum-rules give the  substantially smaller
value $g^{2}\sim 0.08$ $\cite{eletsky}$.

Notice that ${\cal L}^{(1)}$ does not contain the heavy meson masses. This
is the key point  in the implementation of the power counting
of the low energy expansion
as a loop expansion $\cite{J&M}$. The Feynman rules are straightforward to
derive
and the propagators for the heavy mesons  are given by:
\beqa
\Delta_{H}(p)&=&\frac{i}{2\, p.v+i\epsilon}\nn\\
\Delta_{H^{\ast}}^{\mu\nu}(p)&=&-i\;\frac{(g^{\mu\nu}-v^{\mu}v^{\nu})}
{2\, p.v+i\epsilon}~~~~~~~~,
\eeqa
where $p_{\mu}$ is the residual momentum.
The use of these propagators in chiral loop integrals
 is justified, since the physical
cutoff for such integrals is \mbox{ $\Lambda \sim 1 GeV$}. The
implied  change
in the U.V. degree of divergence of the integrals, which occurs at a scale
of momenta of the order of the heavy meson mass,  is therefore irrelevant.
 In calculating the chiral loops it is convenient to use dimensional
regularization, as it preserves chiral invariance. Having eliminated
the heavy meson mass, only low energy scales
appear in the loops, thus furnishing the chiral power counting
as in the case of light mesons $\cite{g&l}$.

The most noticeable effect of \sun\ breaking by the quark masses is in the
masses of the heavy mesons. The leading contribution to the
intramultiplet mass splittings is linear in the light quark masses
( which in chiral power counting are of $\order (p^{2})$ )
and can be described by adding the following
 $\order (p^{2})$ term to the  effective Lagrangian:
\beq
{\cal L}_{\Delta M}^{(2)}=-\frac{C}{4}\; \trd\; \{\;\overline{\Hz} \;(\udag
{\cal M} \udag+
u {\cal M} u)\;\Hz\;\}~~~~~~,
\eeq
where ${\cal M}={\rm diag}(\;m_{u},m_{d},m_{s}\;)$
is the light quark mass matrix$^{\#}$ \footnotetext[1]{{\#}
The form of counterterms containing the quark masses is
determined by implementing full chiral invariance
replacing the quark mass matrix by sources $(s+ip)$
$\cite{g&l}$.}. The \sun\ singlet term
has been omitted as it is of no interest for our purposes.
 Under the assumption that
higher chiral order corrections are small
(more on this in the next section) and
neglecting them,
$C$ is estimated from $M_{D_{s}}-M_{D^{+}}$ and $M_{B_{s}}-M_{B^{0}}$
and the results are: \mbox{$C_{D}=99.5 \pm 0.6 MeV/(m_{s}-m_{d})$},
\mbox{$C_{B}=(82 \pm2.5;121 \pm 10) MeV/(m_{s}-m_{d})$}.
 For  \mbox{$M_{B_{s}}-M_{B^{0}}$}
we use the two values quoted as consistent with present data $\cite{franzini}$.
Establishing this measurement would provide an estimate of
the $1/m_{Q}$ corrections
by comparison of $C_{D}$ with $C_{B}$.
Notice that  \mbox{$C_{K}\sim 225 MeV/(m_{s}-m_{d})$}, as obtained from
isospin breaking in the Kaon masses and using the
ratio $R= (m_{s}-\hat{m})/(m_{d}-m_{u})$ ($\;\hat{m}=(m_{u}+m_{d})/2\;$),
which is substantially larger than
in heavy mesons.
A re-analysis of isospin breaking effects  within the linear approximation
has been recently done $\cite{hou}$. The leading corrections to the linear
approximation are non-analytic in the light quark masses $\sim m^{3/2}$, and
they turn out to be proportional to $g^{2}$, as we will show in next section.
The possibility that these corrections turn out to be important is
not yet  excluded.

\section{Decay Constants and Masses}

In heavy mesons, as in light mesons, the leading
\sun\ breaking correction to  decay constants is proportional
to the light quark masses multiplied by a non-analytic
factor, the chiral logarithm, which emerges due to
the I.R. behaviour of the one chiral loop integrals.

 The decay constants are defined in terms of matrix elements
of the vector and axial vector currents $V_{\mu}^{i}=
\overline{Q} \gmu q^{i}$ and $A_{\mu}^{i}=
\overline{Q} \gmu \gfive q^{i}$ between  one meson state and the vacuum.
In the effective theory, these currents are defined by introducing
vector (${\rm v}_{\mu}$) and pseudovector (${\rm a}_{\mu}$)
 external sources which are triplets under \sun.
These sources couple to the mesons at lowest chiral order ($\order(p)$)
according to the following effective Lagrangian:
\beq
{\cal L}^{(1)}_{{\rm source}}=
\frac{1}{2}\; f\; \trd \;
\{\left({\rm v}_{\mu}^{\dagger}+{\rm a}_{\mu}^{\dagger}\gfive\right)\;
\gmu \;\left[
 (u+\udag)+\gfive (u-\udag)\right]\Hz\} +{\rm h.c.}
\eeq
Clearly, the sources defined here in momentum representation will
only carry residual momentum, and, therefore, they count as quantities
of $\order(p)$.
$f$ is defined in the $m_{Q}\ra \infty$ and chiral limits
 and
related to the usual expressions for the decay constants of
the pseudoscalar and heavy mesons in that limit  by:
\beqa
F_{H}&=&f/\sqrt{M_{H}}\nn \\
F_{H^{\ast}}&=&f \sqrt{M_{H}}=F_{H}M_{H}
\eeqa

The leading \sun\ breaking corrections to the ratios of decay constants and
the leading non-analytic (in the light quark masses) corrections to the mass
differences are determined by calculating the two polarizations \\
\mbox{$\Pi^{A\;ij}_{\mu\nu}(x)=\langle 0\mid T A_{\mu}^{i}(x) A_{\nu}^{j}
(0)\mid 0\rangle $} and
\mbox{$\Pi^{V\;ij}_{\mu\nu}(x)=\langle 0\mid T V_{\mu}^{i}(x) V_{\nu}^{j}
(0)\mid 0 \rangle$}
to one loop order. At long Euclidean distances, the lowest lying
 pseudoscalar
and vector heavy meson poles respectively saturate the two point functions.
These pole contributions  are given in the effective theory
by replacing  the currents  by the effective currents derived from
the source Lagrangian (12). The corresponding Fourier transforms,
which are functions of the residual momentum $p_{\mu}$,  in the
limit $p.v\ra 0$ are given by the following general expressions:
\beqa
\tilde{\Pi}^{A\;ij}_{\mu\nu}(p)&=&\frac{4}{M_{H}}F_{H_{i}}^{2}\;v_{\mu}v_{\nu}
\;\frac{i}{2\,(p.v-\delta M_{i})+i\epsilon}\; \delta_{ij}\nn\\
\tilde{\Pi}^{V\;ij}_{\mu\nu}(p)&=&4M_{H} F_{H^{\ast}_{i}}^{2}
\;\frac{-i\,(g_{\mu\nu}-v_{\mu}v_{\nu})}
{2\,(p.v-\delta M_{i})+i\epsilon}\; \delta_{ij}
\eeqa

The diagrams contributing to one chiral loop order are shown in fig. I.
Since these contributions are U.V. divergent, counterterms are required.
It turns out that one only needs to add counterterms
which  correct the effective currents and which are of
$\order(p^{3})$. Since,  as expected, the chiral loops turn out to preserve
 the heavy quark symmetry, the counterterms
must be invariant under this symmetry as well.
The counterterm effective Lagrangian contains
three low energy constants, and is given by:
\beqa
{\cal L}_{{\rm source}}^{(3)}&=&\nn\\
\!\!\!\!2 B_{0}\;\Gamma_{1} \!\!\!\!\!\!\!
&&\!\!\!\!\!\!\!\frac{f}{F_{0}^{2}} \trd \;
\{\left({\rm v}_{\mu}^{\dagger}+{\rm a}_{\mu}^{\dagger}\gfive\right)
\gamma^{\mu} \left[
 (U {\cal M}u+U^{\dagger}{\cal M}\udag)+\gfive
(U{\cal M} u-U^{\dagger}{\cal M}\udag)\right]\Hz\}\nn\\
+ 2 B_{0}\;\Gamma_{2}\!\!\!\!\!\! \!&&\!\!\!\!\!\!\!
\frac{f}{F_{0}^{2}} \trd\;
\{\left({\rm v}_{\mu}^{\dagger}+{\rm a}_{\mu}^{\dagger}\gfive\right)
\gamma^{\mu} \left[
 ( {\cal M}u+{\cal M}\udag)+\gfive
({\cal M} \udag-{\cal M} u)\right]\Hz\}\\
+ 2 B_{0} \; \Gamma_{3}\; \!\!\!\!\!\!\!&&\!\!\!\!\!\!\!\frac{f}{F_{0}^{2}}Tr
\{ {\cal M} U^{\dagger}+{\cal
M}^{\dagger}U \}
\trd \;
\{\left({\rm v}_{\mu}^{\dagger}+{\rm a}_{\mu}^{\dagger}\gfive\right)
\gamma^{\mu} \left[
 (u+\udag)+\gfive (u-\udag)\right]\Hz\}\nn\\
\!\!\!\!\! +{\rm h.c.}&&~~~~~~~~~~~~~~~~~,\nn
\eeqa
where $B_{0}=\langle \bar{q}q\rangle_{0}/F_{0}^{2}$
is defined in the chiral limit ($M_{\pi}^{2}= B_{0}(m_{u}+m_{d})$, etc.).

We calculate  the loops using dimensional regularization, in which
 it is convenient to
 write: $\Gamma_{j}=\Gamma_{j}^{r}(\mu)+\overline{\Gamma}_{j}
\lambda(\mu),\; (j=1,2,3)$, where $\mu$ is the chiral renormalization
scale, $\Gamma_{j}^{r}(\mu)$ the renormalized effective coupling and
$\lambda(\mu)$ contains the singularity at $d=4$ and is given by:
\beq
\lambda(\mu)=\frac{1}{16\pi^{2}}\mu^{d-4}\{\frac{1}{(d-4)}-\frac{1}{2}
( {\rm log}4\pi +
\Gamma^{\prime}(1)+1)\}\nn
\eeq
 The following choice leads to an U.V. finite result
for the polarizations:
\beqa
\overline{\Gamma}_{1}+\overline{\Gamma}_{2}&=&\frac{5}{24}\;
\left(1+3 g^{2}\right) \nn \\
\overline{\Gamma}_{3}&=&\frac{11}{144}\;
\left(1+3 g^{2}\right)
\eeqa
For the sake of convenience we define: $\Gamma_{12}^{r}(\mu)\equiv
\Gamma_{1}^{r}(\mu)+\Gamma_{2}^{r}(\mu)$.

We first consider \mbox{$\tilde{\Pi}^{A\;ij}_{\mu\nu}$}. One loop contributions
from ${\cal L}^{(1)}$ and ${\cal L}^{(1)}_{{\rm source}}$ and a tree-level
insertion of ${\cal L}^{(2)}_{\Delta M}$ must be included, as shown
  in fig.I (a). The calculation is straightforward, and leads to the
 following results for the pseudoscalar decay constants in the $SU(2)$-limit:
\beqa
F_{H_{u,d}}&=&F_{H}\;\{1-\left(\frac{1+3 g^{2}}{8 F_{0}^{2}}\right)
[3\mu_{\pi}+2\mu_{K}+
\frac{1}{3} \mu_{\eta}]\nn \\
&+&\frac{1}{F_{0}^{2}}\;\Gamma_{12}^{r}(\mu)\;
 2 M_{\pi}^{2}
+\frac{4}{F_{0}^{2}}\; \Gamma_{3}^{r}(\mu)\;\left(2
M_{K}^{2}+M_{\pi}^{2}\right)\}\nn\\
F_{H_{s}}&=&F_{H}\;\{1-\left(\frac{1+3 g^{2}}{8 F_{0}^{2}}\right)
[4\mu_{K}+\frac{4}{3} \mu_{\eta}] \\
&+& \frac{2}{F_{0}^{2}}\;\Gamma_{12}^{r}(\mu) \;
(2 M_{K}^{2}-
M_{\pi}^{2})
+\frac{4}{F_{0}^{2}} \;\Gamma_{3}^{r}(\mu)\; (2 M_{K}^{2}+M_{\pi}^{2})\}
{}~~~~~, \nn
\eeqa
where the chiral logarithms are given by:
\beq
\mu_{P}=M_{P}^{2} \frac{1}{16 \pi^{2}}\; {\rm log}\;\frac{M_{P}^{2}}{\mu^{2}}
\eeq
{}From these expressions one easily finds the \sun\ breaking
corrections to the ratio $F_{H_{s}}/F_{H_{u,d}}$:
\beq
\frac{F_{H_{s}}}{F_{H_{u,d}}}=
1-\left(\frac{1+3 g^{2}}{8 F_{0}^{2}}\right)
[-3\mu_{\pi}+2\mu_{K}+\mu_{\eta}]+
\frac{4}{F_{0}^{2}}\;\Gamma_{12}^{r}(\mu)\;
(M_{K}^{2}-M_{\pi}^{2})
\eeq

\noindent This result, which coincides with that of ref $\cite{Wiseetal}$,
implies that the decay constant grows with the mass of the light quark
in the chiral limit,
as it also occurs in light mesons.
In the case of light mesons, the correction to
$F_{K}/F_{\pi}$ is entirely contained in the
chiral logarithm term if one takes $\mu\sim 1.5\; GeV$ $\cite
{g&l}$. If  a similar situation is assumed
to hold in the case of the heavy mesons, one obtains
 \mbox{$F_{H_{s}}/F_{H_{u,d}}-1\sim 0.13\;(1+3 g^{2})$}
(clearly, this should only be taken as a rough
estimate of the size of the effect). A correction of this size
 should
 be experimentally
 accessible in the future in leptonic $D$-meson decays.  A similar
 direct test is not  available for $B$ mesons
because $B_{s}$ is neutral.

The mass shifts $\delta M_{i}$ receive an $\order (p^{2})$ contribution
from the insertion of ${\cal L}_{\Delta M}^{(2)}$, and a non-analytic
contribution in the quark masses of $\order(p^3)$
 from the loop diagram proportional to
$g^{2}$. The  latter    is similar to the non-analytic
contribution
to the mass splittings in the baryon octet identified long ago
$\cite{pagels}$. The non-analyticity here, as in the case of the
chiral logarithms, is a long distance effect
 due to the I.R. behaviour of the chiral loop integral.
Since local counterterms must be analytic in the quark masses,
these non-analytic corrections are unambiguous, as one would expect
from their long distance nature (for details see $\cite{g&lpr}$).
It is interesting to note that they do not
depend on the fact that we are using the leading term in the
$1/m_{Q}$ expansion within the loop integral; exactly the same
result is obtained by doing the loop integral in the relativistic
theory, and expanding the result.
In the $SU(2)$ limit, we obtain the following expression for the mass
splitting:
\beqa
\delta M_{H_{s}}-\delta M_{H_{u,d}}\left|_{QCD}\right.&=&
 C(m_{s}-\hat{m})-\frac{3 g^{2}}{128 \pi
F_{0}^{2}}[-3 M_{\pi}^{3}+2M_{K}^{3}+M_{\eta}^{3}]\nn\\
\hat{m}=m_{u}=m_{d}
\eeqa
Notice that the non-analytic term has the opposite sign than the leading
term and gives a large
 contribution, unless the effective coupling
$g^{2}$ is very small. For instance, for its contribution
to be less than $20\%$,
$g^{2}<0.06$ is required. For this reason, it is important at some point
to determine this coupling constant with a good degree of confidence,
 since, among
other effects, it could lead to sizeable corrections to the linear
approximation
normally used in the analysis of isospin breaking in heavy mesons.
In particular, $C_{D,B}$ will become closer to $C_{K}$.
A  few remarks are in order here:
a- Due to \sun\ breaking the propagators can
not be brought back to the form (10) because chiral invariance
demands that only \sun\ singlet redefinitions of the form (6) are
admissible. Thus, \sun\ breaking implies that
propagators will in general contain an
$\order(p^{2})$ residual mass. b- The propagator of the heavy meson
in the chiral loop could have been taken with the $\order(p^{2})$
corrections given by the insertion of ${\cal L}^{(2)}_{\Delta M}$.
This, however, only produces a correction of $\order(p^{4})$ while
it does not affect the leading non-analytic term. c- Clearly,
some corrections step by one unit in the chiral power counting,
 making predictions more difficult as they are only suppressed
by a factor $\sim M_{K}/4\pi F_{0}$ which is not much smaller than one.
For instance, it could well
occur that a large leading non-analytic term as that in (21) becomes
partly compensated by a term of $\order (p^{4})$. In this particular
case, an estimate
of the   $\order (p^{4})$ chiral logarithm shows that its contribution
will be small if one chooses $\mu = \order(1 GeV)$, however, the
 $\order (p^{4})$ counterterm could eventually lead to the mentioned
compensation. If $g^{2}$ turns out to be substantially larger
than $0.06$, one would then have an indication for such a compensation.

The relevant one loop contributions to
$\tilde{\Pi}^{V\;ij}_{\mu\nu}(p)$
are shown in fig.I (b), and determine the
corrections to the masses and decay constants of the vector mesons.
 Explicit calculation shows that the
heavy quark symmetry is preserved, as seen in particular
by the relation
$F_{H^{\ast}}=F_{H}M_{H}$, which still holds after the chiral loop corrections
are included.

One might wonder about the precision of the chiral loop corrections
in the heavy quark limit
when applied to $D$ and $B$ mesons.
Calculations of $F_{D}$ and $F_{B}$ in lattice QCD $\cite{fred}$ and QCD
sum rules $\cite{neubert}$ have shown that the $1/\sqrt{m_{Q}}$
scaling characteristic of the heavy
quark limit is not present. On the other hand, there is clear
evidence that the scaling violation
mainly stems from spin independent effects, and therefore, the
heavy quark-symmetry breaking effects
on the ratio
 $F_{H^{\ast}}/F_{H}M_{H}$ are small: ($\sim 10\%$) for the $B$
mesons $\cite{martinelli}$ and somewhat larger for $D$ mesons
 ($\sim 20\%$).
Analogously, the deviation from unity of the ratio
of effective couplings $g_{HH^{\ast}\pi}/g_{H^{\ast}H^{\ast}\pi}$
is expected to be small, since it is also of hyperfine origin.
This deviation and the vector-pseudoscalar mass splitting
are the main source of departure of the chiral corrections
from the $m_{Q}\ra \infty$ limit. We expect, therefore, that
for $D$ and $B$ mesons their departure from this limit will be small
(this involves the reasonable assumption that also the $1/m_{Q}$ corrections
to the counterterms will be small).

\section{$B\ra \overline{D}^{(\ast)}$ Form Factor}

In the infinite mass limit, the heavy quark symmetry permits one
to determine the amplitudes for
$B\ra \overline{D}^{(\ast)}$ transitions mediated by the charged currents
$\;\;\bar{c}\gmu (\gfive) b\;\;$ in terms of a single real form factor
$\xi(v.v^{\prime})$ $\cite{IW1}$, where $v$ ($v^{\prime}$) is the 4-velocity
of the $b$ ($c$) quark. At the vanishing recoil point, $\xi$ satisfies
the normalization condition
\mbox{$\xi(v.v^{\prime}=1)=1$}.

In this section, we will determine the form of the leading $SU(3)$
breaking corrections to this form factor.
To leading order in the chiral and $1/m_{Q}$ expansions the effective
charged currents are obtained from the following source Lagrangian:
\beq
{\cal L}_{b \ra c}^{(1)} = \xi(\nu)
\; \trd \;
\{ \overline{\cal D} \;{\cal B}\;
\left( V^{\mu}+ A^{\mu} \gfive\right)\gmu \} + {\rm h.c.}~~~~~~~,
\eeq
where ${\cal D}$ and ${\cal B}$ are the expressions analogous to
(7) for $D$ and $B$ mesons respectively,
$\nu=v.v^{\prime}$, and  $V^{\mu}$ and  $A^{\mu}$ are
\sun\ singlet  sources. In particular, (22) shows that the
effective  currents do not couple to the Goldstone modes at lowest chiral
order.

The chiral corrections to the matrix element $\langle \overline{D}_{j}\mid
\bar{c}
\gmu b \mid B_{i}\rangle$
are calculated by considering the three-point function\\
\mbox{$\Pi_{\rho\mu\nu}^{V\;i,j}(x,y)=
\langle 0\mid T\;\bar{c} \grho b(0)\;\; \bar{q}_{j} \gnu \gfive c(y)\;\;
\bar{b}
\gmu \gfive q_{i}(x) \mid 0\rangle$}. In the effective meson theory,
 the corresponding three-point
function in residual momentum representation and
 in the limit $p.v\sim p^{\prime}.v^{\prime}
\sim0$, where $p_{\mu}$ ($p_{\mu}^{\prime}$) is the residual momentum
associated with the $B$ ($D$) meson propagator, has the following form:
\beqa
\tilde{\Pi}_{\rho\mu\nu}^{V\;i,j} (p,v,{p'},{v'}) &=&   \\
-\frac{4}{\sqrt{M_{B}M_{D}}}
\;F_{B_{i}}F_{D_{j}} \!\!&v_{\mu} {v'}_{\nu}& \!
\!(v+{v'})_{\rho}\;\xi_{ij}(\nu)\;
\frac{i}{2\, (p.v-\delta M_{i})}\;\frac{i}{2 \,({p'}.{v'}-\delta M_{j})}\nn
\eeqa
To one chiral loop order, the   diagram shown in fig.II (a)
gives the correction
to the I-W form factor, after properly taking into account
 wave function renormalization (the latter naturally emerges when
explicitly calculating all one loop  diagrams for the three-point function).
The result for the form factor is as follows:
\beq
\xi_{ij}(\nu)= \xi(\nu) \left(\delta_{ij}+\omega_{ij}(\nu)\right)\\
\eeq
where $\omega_{ij}(\nu)$ is given by the following expression:
\beqa
\omega_{ij}(\nu)&=& \Omega(\nu)\;
\Delta_{ij}
+ {\rm counterterm}\nn\\
\Delta_{ij}&=& \frac{1}{F_{0}^{2}}\;\sum_{a=1}^{8} (\lambda^{2}_{a})_{ij}
\left( M_{a}^{2} \lambda +\frac{1}{2} \mu_{a}\right)\nn\\
\Omega(\nu)&=&g^{2}
\left( -\frac{3}{2}+(2+\nu)\; A(\nu)+ B(\nu)\right)\\
A(\nu)&=& \frac{1}{2\, \sqrt{\nu^2-1}} \;\;
{\rm log}
\left( \frac{\nu+1+\sqrt{\nu^2-1}}{\nu+1-\sqrt{\nu^2-1}}
\right)\nn\\
B(\nu)&=&\frac{1}{2}+\frac{\nu}{4\,\sqrt{\nu^{2}-1}}\;\;
{\rm log}\left( \frac{\nu-\sqrt{\nu^2-1}}
{\nu+\sqrt{\nu^2-1}} \right)\nn
\eeqa
Clearly,  $\Delta_{ij}$ is diagonal.
Expanding at zero recoil ($\nu=1$), the first few terms are as follows:
\beq
\Omega(\nu)=  g^{2}
\left(-\frac{1}{3}(\nu-1)+\frac{2}{15}(\nu-1)^{2}-
\frac{2}{35}(\nu-1)^{3}+\cdots\right)
\eeq
As expected, from  the fact that the effective vector current
is conserved at leading order in $1/m_{Q}$ as consequence of the
I-W symmetry (more specifically, the part of the symmetry
which corresponds to the flavor rotation between $c$
and $b$ quarks), the corrections vanish
at zero recoil. The counterterms needed to render results U.V. finite
only affect the definition of the effective current.
 They are of $\order (p^{2})$ and
given by:
\beqa
{\cal L}^{(2)}_{b\ra c} \!\!\!\!\!
&=& \!\!\!2 B_{0} \xi(\nu)\frac{ \Omega(\nu)}{F_{0}^{2}}
 \eta_{1}(\nu)\;\trd\; \{\overline{\cal D}
 (u {\cal M} u+\udag {\cal M} \udag) {\cal B}({\rm V}_{\mu} +
{\rm A}_{\mu} \gfive) \gamma^{\mu}\} \\
&+&\!\!\!\!2 B_{0}  \xi(\nu)\frac{ \Omega(\nu)}{F_{0}^{2}}
 \eta_{2}(\nu) \;Tr ({\cal M} U^{\dagger}+{\cal M} U)\;\trd\;
\{\overline{\cal D}\;
  {\cal B}\,({\rm V}_{\mu} + {\rm A}_{\mu} \gfive) \gamma^{\mu}\}\nn
\eeqa
where the following choice of  effective couplings provides a
finite result for the three-point function:
\beqa
\eta_{1}(\nu)&=& \eta_{1}^{r}(\nu;\mu)-
\frac{5}{6}\; \lambda(\mu) \\
\eta_{2}(\nu)&=& \eta_{2}^{r}(\nu;\mu)-\frac{11}{18} \;\lambda(\mu) \nn
\eeqa
Notice that, in accordance with the normalization
condition, the counterterms also have to vanish at zero recoil.

After replacing in $\omega_{ij}$
in (25)  the contribution of the counterterms, one finds the following
\sun\ breaking correction to the ratio of form factors:
\beq
\frac{\large{\xi_{s}(\nu)}}{\large{\xi_{u,d}(\nu)}}=
1+ \frac{\Omega(\nu)}{F_{0}^{2}} \left(
\mu_{K}+\frac{1}{2} \mu_{\eta}-\frac{3}{2} \mu_{\pi} +
4\; \eta_{1}^{r}(\nu;\mu)\; (M_{K}^{2}-M_{\pi}^{2})\right)
\eeq
This result shows that for small quark masses the chiral logarithm
tends to increase the value of $\xi_{s}$ with respect to $\xi_{u,d}$.
This result seems to be anti-intuitive; one would expect that the
heavier the light quark is the faster the form factor will drop with
increasing $\nu$. This is certainly true for large enough light quark mass.
As the chiral limit is approached, however, the behavior reverses.
Such a behaviour is known to occur, for instance, in the quark-antiquark
condensates, which, near the chiral limit increase with the quark
mass, yet start decreasing as the mass becomes large enough.
Notice that the form of the correction is algebraically of the same
form as the one for the ratio of decay constants. The explicit $\nu$-dependence
of $\eta_{12}$ indicates that the counterterm can change the profile
of the correction with respect to that  given by $\Omega(\nu)$.

It is important to emphasize
 that the chiral loop result (24, 25) holds for any value of the
recoil. The behavior of $\Omega(\nu)$ is smooth, and
for \mbox{$\nu \ra \infty$} it tends to $\;-g^{2}$.
 In particular, it applies to the whole Dalitz domain of the semileptonic
decays
$B\ra \overline{D} l\nu_{l}$, and to non-leptonic decays in the
factorization limit (e.g.
$B^{0}\ra {D}^{+} \pi^{-}$ and $B_{s}\ra \overline{D}_{s} \pi^{-}$).
At the largest available recoil $\nu\sim 1.8$, we have
 $\Omega \sim -0.2\; g^{2}$.
Choosing values for $\mu$ between $1$ and $1.5\;\;GeV$, the correction to
the the ratio (29) contributed by the chiral logarithms is
$\sim -\Omega(\nu)\times(0.3 \;{\rm to}\; 0.5)$. This implies that even
at the largest recoil this contribution will be small, perhaps, only a
few percent. Thus, unless the counterterm contribution
is surprisingly large, the $SU(3)$ breaking effects
on the form factor will be in the few percent range.
Notice that the size of this correction and the
non-analytic contributions to the mass splittings are related, since
both are proportional to $g^{2}$.

As in the case of the decay constants, we explicitly checked that the
heavy quark symmetry is preserved by the chiral loops. This check was
done by considering the three-point function \\
\mbox{$\Pi_{\rho\mu\nu}^{A\;i,j}(x,y)=
\langle 0\mid T\;\bar{c} \grho \gfive b(0)\;\; \bar{q}_{j} \gnu  c(y)\;\;
 \bar{b}
\gmu  q_{i}(x) \mid 0\rangle$}.
In this case, the corrections to the form factor are
 obtained from the diagrams  in fig.II (b).

Finally, \sun\ breaking gives rise to direct couplings of the Goldstone
bosons to the effective currents, as shown by the counterterm (27). They are
proportional to the light quark masses, and therefore, small and
unlikely to be of direct physical relevance.

\section{Remarks}

It is certainly worthwhile to explore  possible effects of
chiral symmetry in the physics of heavy hadrons. It
is likely that they  will mainly be restricted
to the light quark mass induced \sun\ breaking corrections as
the ones discussed in this work. Less clear is the accessibility
to the predictions of low energy theorems for the
soft meson emission in decays. At any rate, substantial experimental
improvement, especially in strange heavy mesons, is required
until  effects beyond the mass splittings are accessible.
As the prediction power of the chiral expansion is limited
by the need of introducing counterterms (as we saw in the cases
of the decay constants and the I-W form factor), it is not clear
that enough experimental information will  become available
 as to reach the stage of testing these limited predictions.

We expect that $F_{D_{s}}/F_{D}$, for which the \sun\ breaking
correction might be of the order of $10 - 20 \;\%$, to be a first
candidate to be measured. The corrections to the I-W form factor
will be much harder to observe, since this will  require
large numbers of $B_{s}$ mesons, and the effect itself
might be very small. The situation could be  improved by
considering some semi-inclusive decays where the chiral expansion
is well defined, for instance, channels for which factorization
holds to a good degree. The leading non-analytic corrections
to the mass splittings might be large, depending on the value
of $g^{2}$, and could, therefore, affect analyses on isospin
breaking done within the linear approximation.

Lattice QCD simulations of heavy hadrons might also be an
interesting domain of application. For example, chiral symmetry
 controls the finite volume effects, which are fully
predictable for the unquenched theory. The problem of determining
the effective couplings, like those appearing in
the counterterms ${\cal L}_{\rm source}^{(3)}$ and
${\cal L}_{b\ra c}^{(2)}$ might well be first
solved on the lattice, by looking at the light quark-mass
dependence of the observables we discussed.

\vspace{0.5 cm }
I thank H. Leutwyler for
enlightening remarks on non-analytic terms and
  S. Sharpe for informative comments and for making me aware of
the recent works on the topic of this paper.
I also appreciate the support from the Paul Scherrer Institute, where
part of this work was done.

\newpage

{\large{Figure Captions}}\vspace*{2cm}

\noindent Fig. I : One loop contributions to the polarizations a)
$\tilde{\Pi}^{A\;ij}_{\mu\nu}$
and b) $\tilde{\Pi}^{V\;ij}_{\mu\nu}$. The solid lines correspond to the heavy
pseudoscalar, the wavy line to the heavy vector and the dashed line
to the Goldstone bosons. The square dot represents the insertion of
${\cal L}_{\Delta M}^{(2)}$. Diagrams not explicitly shown vanish
identically.\vspace*{1cm}

\noindent Fig. II : One loop correction to the form factor
$\xi(v.v^{\prime})$ as determined from the three point functions
a) $\tilde{\Pi}_{\rho\mu\nu}^{V\;i,j}$
and b) $\tilde{\Pi}_{\rho\mu\nu}^{A\;i,j}$.
 Diagrams which only contribute
to decay constants, masses and wave function renormalization
are not displayed. \vspace*{1cm}

\end{document}